\documentclass[aps,pra,twocolumn,groupedaddress,amsmath,amssymb,nofootinbib]{revtex4-1}
\usepackage{graphics}
\usepackage{natbib}

\newcommand{\Tr}{\text{Tr}}
\newcommand{\Cov}{\text{Cov}}
\newcommand{\be}{\begin{equation}}
\newcommand{\ee}{\end{equation}}
\newcommand{\bea}{\begin{eqnarray}}
\newcommand{\eea}{\end{eqnarray}}
\newcommand{\ket}[1]{\left|#1\right\rangle}

\newcommand{\expected}[1]{\left\langle#1\right\rangle}
\newcommand{\var}[1]{\expected{\Delta^2#1}}

\begin{document}
\title{Transmission Estimation at the Cram\'er-Rao Bound for Squeezed States of Light in the Presence of Loss and Imperfect Detection}
\author{Timothy S. Woodworth$^{1,2}$, Kam Wai Clifford Chan$^{3}$, Carla Hermann-Avigliano$^{4,5}$, and Alberto M. Marino$^{1,2}${\footnote {marino@ou.edu}}}
\affiliation{$^{1}$Homer L. Dodge Department of Physics and Astronomy, The University of Oklahoma, Norman, Oklahoma, 73019, USA}
\affiliation{$^{2}$Center for Quantum Research and Technology, The University of Oklahoma, Norman, Oklahoma, 73019, USA}
\affiliation{$^{3}$OAM Photonics LLC, San Diego, CA 92126, USA}
\affiliation{$^{4}$Departamento de Física, Facultad de Ciencias Físicas y Matemáticas, Universidad de Chile, Santiago, Chile}
\affiliation{$^{5}$Millennium Institute for Research in Optics (MIRO), Concepción, Chile}

\begin{abstract}
Enhancing the precision of a measurement requires maximizing the information that can be gained about the quantity of interest from probing a system. For optical based measurements, such an enhancement can be achieved through two approaches, increasing the number of photons used to interrogate the system and using quantum states of light to increase the amount of quantum Fisher information gained per photon. Here we consider the use of quantum states of light with a large number of photons, namely the bright single-mode and two-mode squeezed states, that take advantage of both of these approaches for the problem of transmission estimation. We show that, in the limit of large squeezing, these states approach the maximum possible quantum Fisher information per photon for transmission estimation that is achieved with the Fock state and the vacuum two-mode squeezed state. Since the bright states we consider can be generated at much higher powers than the quantum states that achieve the maximum quantum Fisher information per photon, they can achieve an much higher absolute precision as quantified by the quantum Cram\'er-Rao bound. We discuss the effects of losses external to the system on the precision of transmission estimation and identify simple measurements techniques that can saturate the quantum Cram\'er-Rao bound for the bright squeezed states even in the presence of such external losses.
\end{abstract}
\maketitle

\section{Introduction}

There is currently a significant interest in taking advantage of quantum resources for applications that range from quantum enhanced sensing~\cite{Olson2017,Zhang2013,Kacprowicz2010,Dowran2018,Degen2017} to quantum simulations~\cite{Cirac2012,Hartmann2016} to quantum communications~\cite{Gisin2007,Orieux2016,Furrer2018,Zhang2018}. Among these applications, quantum enhanced sensing is the most likely to find its way in to real-life devices in the near future. In particular, optical based quantum enhanced sensing, for which quantum states of light play a very prominent role~\cite{Lawrie19}, has already led to significant enhancements for gravitational wave detection. Advanced LIGO now uses a vacuum single-mode squeezed state injected into the unused port of the interferometer to decrease the noise below the standard quantum limit~\cite{Aasi2013}.

While most of the work on quantum enhanced sensing with light has focused on phase estimation~\cite{Dorner2009,Aasi2013,Seshadreesan2013,Zhang2017}, as is the case with LIGO, the promise of quantum enhanced devices and measurements has recently led to an increased interest in the estimation of other parameters, such as transmission through an optical system~\cite{Scheel2003,Monras2007,Adesso2009,Invernizzi2011,Moreau2017, Losero2018}. In performing these studies, the quantum Cram\'er-Rao bound (QCRB)~\cite{Helstrom1976, Holevo1982, PARIS2009}, which sets the limit to the precision that can be achieved in the estimation of a parameter, provides an absolute metric that can be used to compare sensitivity limits when proving the system of interest with different states, either quantum or classical.

Here, we focus on the problem of transmission estimation as it is at the heart of a number of applications. For example, enhanced transmission estimation can tighten theoretical bounds for quantum-enhanced measurement schemes~\cite{Demkowicz-Dobrzanski2009,Kacprowicz2010,Zhang2013}, can enable precise calibration of photodiodes and other optical systems, can set limits to key rate in quantum communications~\cite{Qi2019,Fang2020}, and can enhance sensors whose operation is dependent on a change in transmission~\cite{Dowran2018}.  For this problem, it has been shown that the Fock state~\cite{Adesso2009} and the vacuum two-mode squeezed state (vTMSS)~\cite{Invernizzi2011} give the ultimate limit in sensitivity for a given number of probing photons and that the vacuum single-mode squeezed state approaches this limit for large transmissions~\cite{Monras2007}. However, these states can only be generated with a low number of photons, which limits their ability to surpass the classical state-of-the-art as the sensitivity increases with number of probing photons. Thus, classical states of light, which can easily be generated with high powers, can achieve a higher absolute sensitivity (or lower QCRB) than the optimal quantum states.

In order to overcome the limitations of the optimal quantum states, we propose the use of quantum states that can be generated with a macroscopic number of photons, such as bright single-mode squeezed states (bSMSS) and bright two-mode squeezed states (bTMSS). We show that these states have a sensitivity for a given number of probing photons that is much higher than the one for a classical coherent state and approaches that of the Fock state and vTMSS in the limit of large squeezing. Thus, these states can lead to an enhancement both due to their large number of photons and their quantum properties, giving an overall lower QCRB. While the QCRB provides the limit for precision in the estimation of a parameter, the problem of finding a measurement that saturates it is not a trivial one and is not always possible. We show that optimized intensity measurements saturate the QCRB for both the bSMSS and bTMSS. Furthermore,  we show that these schemes still saturate the QCRB even in the presence of losses external to the system under study.  These results provide a way to surpass the classical-state-of-the-art for real-life practical applications where there is a limit to the maximum power that can be used to probe a system.

\section{Quantum Cram\'er-Rao Bounds}

\begin{figure}[h]
	\includegraphics{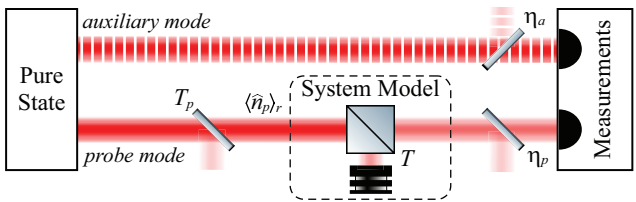}
	\caption{Configuration for the estimation of the unknown transmission $T$ of a system. We use a beamsplitter to model the transmission through the system and consider the use of both single- and two-mode states of light. One of the modes is used to probe the system, while for the case of a two-mode state the second mode serves as an auxiliary mode. We consider losses external to the system  for the probe with a transmission $T_p$ before and $\eta_p$ after the system.  The auxiliary mode is subject to a transmission $\eta_a$ before detection.\label{fig:BlockDiag}}
\end{figure}

The problem we consider, shown schematically in Fig.~\ref{fig:BlockDiag}, is the use of either a single- or a two-mode state of light to estimate the unknown transmission $T$ of a system. In the proposed configuration one of the modes is used to directly probe the system. When the state consists of two modes, the second mode is used as an auxiliary mode that can serve as a reference. After interacting with the system, measurements on the mode(s) are performed to estimate the transmission. We take the number of photons used to probe the system, i.e. mean number of photons in the probe mode right before the system ($\expected{\hat{n}_{p}}_{r}$), as the resource for the parameter estimation problem. We therefore compare the transmission estimation QCRB for different states with the same  number of photons probing the system. To consider more realistic operational conditions, we study the effects of loss external to the system both before the system, to allow an extension to mixed states, and after the system, to account for detection inefficiencies, as well as losses on the auxiliary mode.

To model the system we use a beamsplitter (BS), which is the most often used model for linear transmission~\cite{Kim2002}. As a model for transmission, one input port of the BS couples in the probe mode while the other port, which is assumed to be unaccessible, couples in the vacuum. At the output, one port corresponds to the transmitted portion of the probe mode while the other port corresponds to the portion of the probe mode that is lost to the environment and is undetectable. The unitary operation performed by the BS is given by the operator~\cite{Kim2002}
\be
\hat{B}=e^{\cos^{-1}(\sqrt{T})\left(\hat a_{p}^\dagger\hat a_\nu^{}-\hat a_{p}^{}\hat a_\nu^\dagger\right)},
\ee
where $\hat a_{p}~(\hat a_\nu)$ is the annihilation operator of the probe (vacuum) mode and $T$ is the intensity transmission of the system. Without loss of generality, the amplitude transmission is taken to be real as the parameter of interest is the intensity transmission and not the phase. The same BS model is used to take into account losses external to the system.

In estimating a parameter such as transmission, the Cram\'er-Rao bound~\cite{Helstrom1976, Holevo1982, PARIS2009}, given by the inverse of the Fisher information $F(T)$
\be
\expected{\Delta^2 T}\ge\frac{1}{F(T)},\label{eqn:classicalCRB}
\ee
provides a lower bound for the estimation uncertainty of the parameter of interest ($\expected{\Delta^2 T}$). In defining Eq.~(\ref{eqn:classicalCRB}), there is a bound given by the inverse of the classical Fisher information, $F_C(T)$, which is a metric for distinguishability of the distribution of measurement results for different values of the parameter.  Thus, the so called classical Cram\'er-Rao bound is dependent on the measurement apparatus and is based on measurement results. The classical Cram\'er-Rao bound can be generalized by optimizing over all possible measurements to maximize the classical Fisher information. This approach leads to the so called quantum Fisher information (QFI), $F_Q(T)$, which is a metric for distinguishability of the density matrix of the state after interacting with the system for different values of the parameter. Therefore, in accordance with Eq.~(\ref{eqn:classicalCRB}), the QCRB states that the lowest variance in the estimation of a parameter for a given state and system is given by the inverse of the QFI.

It is also possible to calculate an upper bound for the QFI that can serve to identify the optimal quantum states  and thus provide an ``ultimate'' bound for the estimation of transmission. Following the method introduced by Escher \emph{et al.}~\cite{Escher2011}, we assume that the portion of the probing mode that leaves the BS through the output port that is coupled to the environment can be measured and therefore no information, encoded in the number of photons, is lost. While non-physical, this assumption sets an upper limit for realistic QFIs. Measuring the number of photons leaving both output ports of the BS, we find that the maximum QFI, $\mathcal{F}_\text{max}(T)$, for any state is given by
\be
F_C(T)\le F_Q(T)\le\mathcal{F}_\text{max}(T)=\frac{\expected{\hat n_p}_r}{T-T^2},\label{eq:MaxFisher}
\ee
which is equal to the result found in~\cite{Monras2007}.  When $F_Q=\mathcal{F}_\text{max}$ the probing state acquires the maximum QFI of any state and when $F_C=F_Q$ the implemented measurement strategy is able to extract the maximum information about the parameter of interest and represents the optimal measurement strategy. Thus, when all three quantities are equal, the ultimate sensitivity limit in the estimation of $T$ is achieved for a given number of probing photons $\expected{\hat n_p}_r$.

Since transmission estimation scales as the inverse of the number of probing photons, we can define an estimation function, $\Lambda(T)=\expected{\Delta^2 T}\expected{\hat n_p}_r$, as the inverse of the QFI per photon or, equivalently, the QCRB multiplied by the number of probing photons. This makes it possible to discuss the sensitivity that can be obtained when probing a system with a certain state independent of the number of photons used to probe it, while still discussing the QCRB that takes into account the number of probing photons. Below, we show that  bright squeezed states have estimation functions that approach the one of the vTMSS and Fock state in the limit of large squeezing.

\subsection{Estimation Function for Pure States}

It has been previously shown that both the Fock state~\cite{Adesso2009} and the vTMSS~\cite{Invernizzi2011} can saturate the maximum limit for the QFI, $\mathcal{F}_\text{max}(T)$, and as a result are the optimal quantum states for transmission estimation for a given number of photons. Their estimation functions are given by
\be
\Lambda^{\rm vTMSS}=\Lambda^{\rm Fock}=\Lambda_{\rm min}=T-T^2.
\ee
For the Fock state, this result can be understood given that probing with exactly $N$ photons and measuring $N-N_e$ photons on the output implies that exactly $N_e$ photons were lost to the environment. While $N_e$  is different for each realization, the information lost to the environment can be recovered from the transmission due to perfect knowledge of the number of incident photons and the relation between the two output ports of the BS from energy conservation.
 For the vTMSS, the fact that photons in the two modes are always present in pairs makes it such that any photon lost in the probe mode will have its pair photon still present in the auxiliary mode. As such, the number of photons lost to the environment is the same as the photon number difference between the probe and auxiliary modes. In fact, this property has been previously proposed and used for the absolute calibration of the quantum efficiency of photon counting detectors and  photodiodes~\cite{Worsley2009,Marino2011,Cohen2018}.

While the Fock state and the vTMSS have the lowest possible estimation function, currently they can only be generated with a low mean photon number~\cite{Agafonov2010,Chekhova2015}. Thus, in practice, even coherent states can reach a lower QCRB due to the fact that they can be generated with a very large number of photons that can be used to probe the system.  To overcome this limitation of the Fock state and vTMSS, we consider the use of bright squeezed states, both the bTMSS and bSMSS, which can be generated with significantly higher mean number of photons by seeding the parametric process used to generate them.

We first consider the bTMSS for the case in which the two modes are seeded with coherent states of complex amplitude $\alpha$ for the probe and $\beta$ for the auxiliary mode. We calculate its QFI using the method outlined by {\v{S}}afr{\'{a}}nek et al.~\cite{Safranek2015} and  show in Appendix~\ref{app:PureTMSS} that it is given by
\be
F_Q^{\rm bTMSS}=\frac{\expected{\hat n_p}^\text{vac}}{T-T^2}+\frac{\expected{\hat n_p}^\text{bright}}{T-T^2+T^2\text{sech}(2s)},\label{eqn:FullTMSSQFI}
\ee
where $\expected{\hat n_p}^\text{vac}=\sinh^2(s)$ is the number of spontaneously generated photons and $\expected{\hat n_p}^\text{bright}=|\alpha|^2\cosh^2(s)+|\beta|^2\sinh^2(s)-|\alpha||\beta|\cos(\Theta)\sinh(2s)$ is the number of stimulated photons in the probe mode. The $s$ parameter~\cite{Drummond2003,McCormick2008} characterizes the rate of photon pair generation, $\Theta$ takes into account the phases of the squeezing process and input seed coherent states, and $|\alpha|^2$ ($|\beta|^2$) is the mean number of seed photons for the probe (auxiliary) mode. Through out the paper, we use the approach introduced by Yuen~\cite{Yuen1976} of applying the squeezing unitary on the input seed coherent states. Such an approach more closely resembles the experimental approach of using coherent states to seed the parametric process that generates the bright squeezed states.

\begin{figure}
	\includegraphics{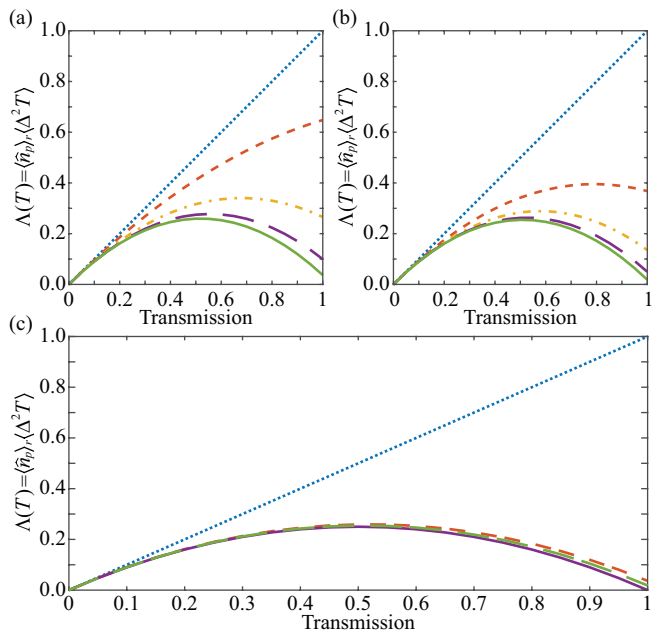}
	\caption{Estimation function as a function of transmission for the (a) bTMSS and (b) bSMSS for different values of $s$: 0 (blue dotted), 0.5 (red short dashed), 1.0 (yellow dot-dashed), 1.5 (purple long dashed), and 2.0 (green solid). As can be seen, the bSMSS estimation function is always lower than the one for the bTMSS for the same value of $s$. (c) Comparison of the estimation functions for the coherent state (blue dotted), bTMSS (red short dashed), bSMSS (green long dashed), and Fock/vTMSS (purple solid) for an $s$ value of 2 for the squeezed states. The bright squeezed state estimation functions are close to the ultimate limit established by the Fock and vTMSS and much lower than the one for the coherent state. \label{fig:PureStates}}
\end{figure}

In the bright seed limit in which the stimulated term dominates the QFI we can drop the first term on the right hand side of Eq.~\eqref{eqn:FullTMSSQFI}, such that the bTMSS estimation function takes the form
\be
\Lambda^{\rm bTMSS}=T-T^2\left[1-\text{sech}(2s)\right].\label{eqn:bTMSSQFI}
\ee
In this limit, the number of probing photons are given by $\expected{\hat n_p}_r\cong\expected{\hat n_p}^{\rm bright}$, as we can ignore the contribution of the spontaneously generated photons. As can be seen from Eq.~\eqref{eqn:bTMSSQFI}, as $s$ increases
$\Lambda^{\rm bTMSS}\rightarrow\Lambda_\text{min}$, which means that for the same number of photons the bTMSS tends towards the ultimate limit given by the Fock state, as shown in Fig.~\ref{fig:PureStates}(a). With experimentally realizable levels of squeezing, with $s\approx2$~\cite{Vahlbruch2016}, the bTMSS does not reach the same estimation function as the Fock state and vTMSS, though the difference between the estimation functions is small, as seen in Fig.~\ref{fig:PureStates}(c).
The bTMSS, however, can be generated with significantly higher powers than either the Fock state or vTMSS, with $\sim$1~mW of power for the probing beam~\cite{McCormick2008}. This translates to a photon flux of $\sim\!\!10^{15}$ photons/second at 800~nm, a level Fock states~\cite{Uria2020} and even bright vTMSS~\cite{Agafonov2010,Chekhova2015} cannot reach. Thus, if we take into account the maximum number of probing photons that can be generated in practice, the bTMSS has a QCRB many orders of magnitude smaller that either the Fock state or vTMSS.

Next, we consider the bSMSS, for which all the photons generated by the source are used to probe the system. We again use the technique from {\v{S}}afr{\'{a}}nek et al. and follow a procedure analogous to the one shown in Appendix~\ref{app:PureTMSS} to show that the estimation function takes the form
\be
\Lambda^{\rm bSMSS}=T-T^2\left(1-e^{-2s}\right)
\ee
when the system is probed with an amplitude bSMSS. This equation matches the results from reference~\cite{Monras2007}, which treats the general case of a single-mode Gaussian state.

As can be seen in Fig.~\ref{fig:PureStates}, the bSMSS has a lower estimation function than the bTMSS. Thus, in theory the bSMSS would be a better option for transmission estimation, as it makes better use of quantum resources. In practice, however, the presence of an auxiliary mode makes it possible to perform differential measurements that can cancel classical technical noise that would otherwise limit or completely inhibit any quantum enhancement.  Therefore, careful considerations of the experimental limitations need to be made when deciding between the use of a bSMSS or a bTMSS.

In the limit in which the squeezing parameter, $s$, tends to zero the estimation functions for the bTMSS and bSMSS become the same as the one for the coherent state, as would be expected, that is
\be
\left.\begin{matrix}\Lambda^{\rm bTMSS}\\\Lambda^{\rm bSMSS}\end{matrix}\right\}\underset{s\rightarrow0}{\Rightarrow}\Lambda^{\rm Coh}=T.
\ee
Thus, the QCRB for the coherent state scales linearly with transmission in the same way as the intensity variance of this shot-noise limited state.

As can be seen in Fig.~\ref{fig:PureStates}, both the bTMSS and bSMSS are able to obtain a very significant quantum enhancement for the same number of probing photons for large transmission in the limit of large squeezing.  As the transmission of the system decreases, the estimation function of all quantum states tend to the one of the coherent state.  This is to be expected, as any state tends to a coherent state in the presence of losses due to the vacuum that couples in through one of the input ports of the BS.  Thus, in the limit of low transmission, the noise will be dominated by the vacuum noise.

\subsection{QCRB for Mixed States}

For realistic operational conditions, it is difficult to generate a pure optical quantum state,  propagate it without loss to the system, and then perfectly detect it. It is therefore important to consider the impact of these imperfections. Here, we consider the case in which it is possible to generate a pure state and any mixedness of the state results from  a loss mechanism that can be modeled as a BS.

To take into account such sources of losses external to the system of interest, the probe mode is transmitted through a BS before and after the system with intensity transmission $T_p$ and $\eta_p$, respectively, while the auxiliary mode is transmitted through a BS with intensity  transmission $\eta_a$, as shown in Fig.~\ref{fig:BlockDiag}.  In the presence of these external losses, we still consider the resources to be given by the number of photons incident on the system of interest, such that $\expected{\hat n_p}_r=T_p\expected{\hat n_p}_0$, where $\expected{\hat n_p}_0$ is the number of probe photons generated by the source. The losses after the system for the probe and total losses for the auxiliary mode make it possible to take into account additional optical losses as well as the effect of an imperfect detector, both of which will lead to the loss of information.

We show in Appendix~\ref{app:Loss} that once we take these loss mechanisms into account, the estimation functions take the form
\bea
\Lambda^{\rm Coh}&=&\frac{T}{\eta_p}\label{eq:coherent_losses}\\
\Lambda^{\rm bTMSS}&=&\frac{T}{\eta_p}-T^2T_pH_a\left[1-\text{sech}(2s)\right]\label{eq:LossyBTMSS}\\
\Lambda^{\rm bSMSS}&=&\frac{T}{\eta_p}-T^2T_p\left(1-e^{-2s}\right)\\
\Lambda^{\rm Fock}&=&\frac{T}{\eta_p}-T^2T_p,
\eea
where $H_a$ is given by
\be
H_a=\frac{(2\eta_a-1)\left[1+2\sinh^2(s)\right]}{1+2\eta_a\sinh^2(s)}.
\ee
Note that the effects of losses for the auxiliary mode are all contained in $H_a$, such that when the auxiliary mode is perfectly detected, $\eta_a=1$, then $H_a=1$.  At 50\% loss of the auxiliary mode, $H_a=0$ and the bTMSS estimation function is the same as the coherent state estimation function. As the loss increases above 50\%, $H_a$ becomes negative and second term on the right hand side changes sign, which leads to an estimation function larger than the one for a coherent state. Thus, low loss of the auxiliary mode is needed in order to obtain a quantum enhancement with the bTMSS. As can be seen from these equations, while the bTMSS estimation function has the same relation to losses of the probing beam as the single-mode states, it is further degraded  by losses of the auxiliary mode through $H_{a}$. This, for example, makes the bTMSS more susceptible to detection efficiencies than the bSMSS.

\begin{figure}
	\includegraphics{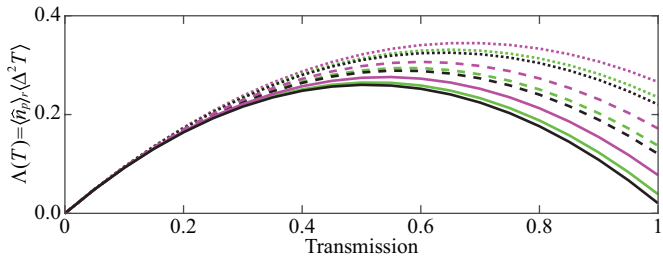}
	\caption{Effect of losses before the system on the estimation function for the Fock state (black), bSMSS (green), and bTMSS (magenta) with $\eta_a=\eta_p=0.98$.  We consider $T_p$ of 1 (solid), 0.90 (dash), and 0.80 (dotted). A squeezing parameter of $s=2$ is used for the squeezed states. \label{fig:LossySQZstates}}
\end{figure}

The linear term on the right hand side of the estimation functions for the states considered is the same for all of them and is the only term affected by probe losses after the system. As can be seen from Eq.~(\ref{eq:coherent_losses}), this term is the classical limit for estimating transmission. When photons are lost after the system the QFI they carry is also lost, which leads to an increased variance in the transmission estimation. As can be seen from the equations above, the quadratic term depends on the intensity noise or intensity correlations of the overall state that is considered.

To study the effect of using a mixed state to probe the system, in Fig.~\ref{fig:LossySQZstates} we compare the estimation functions for the Fock state, bSMSS, and bTMSS for different losses on the probe before interacting with the system, $T_{p}$. We consider the case in which the only loss after probing the system and for the auxiliary mode is due to the quantum efficiency of the photodiodes, which is taken to be the same for both of them.  The effect of losses before the system on the estimation function is more significant for higher transmissions of the system, $T$, due to the larger quantum enhancement that can be achieved in this limit. This is a result of the quadratic term, which gives the quantum enhancement, decreasing linearly with pre-system transmission, $T_p$.  This can be understood by considering that the intensity variance of the quantum states tends towards a Poissonian distribution, same as the statistics of a coherent state, as losses increase. Thus, as the transmission before the sample tends to zero, the state tends towards a coherent state and the quadratic term tends to zero.

\section{Measurements that Saturate the QCRB}

While the QCRB provides a fundamental limit to the sensitivity that can be achieved for a given quantum state, it does not provide a means to identify a measurement that can saturate it. Finding such measurements is non-trivial and, in general, may not be physically implementable. Here we show that for the bright squeezed states there exists a physical measurement that saturates the QCRB, one that can be easily implemented with current technology.

\begin{figure}
	\includegraphics{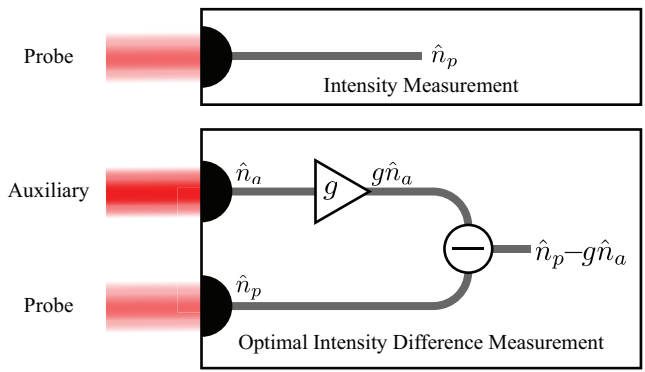}
	\caption{Measurements that saturate the QCRB for the coherent, bSMSS, and Fock state (top) and the bTMSS (bottom). The top diagram corresponds to an intensity measurement while the bottom one corresponds to the optimal intensity difference measurement. The optimal intensity difference measurement has electronic gain on the auxiliary mode detection as a parameter to maximize the cancelation of the probe mode fluctuations after subtraction of the two modes. \label{fig:IntDiff}}
\end{figure}

We show that the Fock, intensity squeezed bSMSS, and coherent state QCRBs can all be saturated by an intensity measurement, consistent with previous results for photon counting with Fock states~\cite{Adesso2009}. This measurement strategy, shown in the top diagram of Fig.~\ref{fig:IntDiff}, consists of a photodiode (or photon counting detector for Fock states) used to measure the number of photons in the probe mode. As shown in Appendix~\ref{app:Measure}, when both the system and external losses are taken into account, the variance in intensity (photon number) with such a measurement strategy takes the form
\be
\expected{\Delta^2 \hat n_p}=(TT_p\eta_p)^2\expected{\Delta^2\hat n_p}_0+TT_p\eta_p(1-TT_p\eta_p)\expected{\hat n_p}_0\label{eq:IntensityMeasure}
\ee
where $\var{\hat n_p}_0$ and $\expected{\hat n_p}_0$ are the variance and mean for the number of photons at the output of source, respectively. As expected, as the total transmission decreases, the states tend towards a coherent states for which the variances is equal to the mean.  The uncertainty in transmission estimation for the intensity measurement can then be obtained through error propagation to be given by
\be
\var{T}=\frac{T}{\eta_p\expected{\hat n_p}_r}-\frac{T^2T_p}{\expected{\hat n_p}_r}\left(1-\frac{\var{\hat n_p}_0}{\expected{\hat n_p}_0}\right),\label{eqn:singleModeMeasure}
\ee
where
\be
\frac{\var{\hat n_p}_0}{\expected{\hat n_p}_0}=\begin{cases}
			1&\text{coherent state}\\
			e^{-2s}&\text{bSMSS}\\
			0&\text{Fock state}
			\end{cases}\label{eqn:Fanos}
\ee
is the Fano factor for each state. As can be seen by these results, intensity measurements can saturate the transmission estimation QCRB for coherent states, bSMSS, and Fock states even in the precesses of losses external to the system of interest.

We next consider the case of the two-mode state, the bTMSS.  We show that for this state the transmission estimation QCRB can be saturated by optimized intensity difference measurements, $\hat n_-^g=\hat n_p-g\hat n_a$, shown on the bottom diagram of Fig.~\ref{fig:IntDiff}.  For this measurement strategy an electronic gain, $g$, is used after the auxiliary mode photodiode to minimize the measured noise by optimizing the  cancellation of the probe mode power fluctuations. While in principle it would be possible to instead optically attenuate the auxiliary mode, as it effectively scales the measured fluctuations of the auxiliary mode, it would further reduce the correlations between the probe and auxiliary modes. Optical loss would randomly remove photons from the auxiliary mode, some of which could be correlated to photons in the probe mode. On the other hand, adjusting the gain electronically does not lead to additional loss of correlations; however, it still makes it possible to scale the measured auxiliary mode fluctuations to minimize the intensity difference noise. As we show, the optimized intensity difference can always saturate the bound for the case in which the source used to generate the bTMSS is seeded only with the probe beam ($\alpha\neq0$, $\beta=0$). For the case in which both modes are seeded ($\alpha\neq0$, $\beta\neq0$) the bound is only saturated for a phase of the parametric process that minimizes the variance of the optimized intensity difference measurement, that is $\cos(\Theta)=-1$.

Following the same procedure as the one used in Appendix~\ref{app:Measure}, we shown that once the system and external losses are taken into account, the intensity difference measurement can be written as
\bea
\var{\hat n_-^g}&=&\var{\hat n_p}+g^2\var{\hat n_a}-2g\,\text{Cov}(\hat n_p,\hat n_a)\\
&=&(TT_p\eta_p)^2\expected{\Delta^2\hat n_p}_0+TT_p\eta_p(1-TT_p\eta_p)\expected{\hat n_p}_0\notag\\
&&+g^2\left[\eta_a^2\expected{\Delta^2\hat n_a}_0+\eta_a(1-\eta_a)\expected{\hat n_a}_0\right]\notag\\
&&-2gTT_p\eta_p\eta_a\big(\expected{\hat n_p\hat n_a}_0-\expected{\hat n_p}_0\expected{\hat n_a}_0\big),\label{eqn:OptIntDiff}
\eea
where Cov$(\hat n_p,\hat n_a)$ is the covariance between the photon numbers in the probe and auxiliary modes and the subscript $0$ indicates the generated state. Equation~(\ref{eqn:OptIntDiff}) is minimized when the electronic gain is set to
\be
g_{\rm opt}=TT_p\eta_p\frac{\expected{\hat n_p\hat n_a}_0-\expected{\hat n_p}_0\expected{\hat n_a}_0}{\eta_a\expected{\Delta^2\hat n_a}_0+(1-\eta_a)\expected{\hat n_a}_0}.
\ee
The mean and variance of the optimized intensity difference measurement are then given by
\bea
\expected{\hat n_-^{g_{\rm opt}}}&=&TT_p\eta_p\expected{\hat n_p}_0-g_{\rm opt}\eta_a\expected{\hat n_a}_0\\
\var{\hat n_-^{g_{\rm opt}}}&=&(TT_p\eta_p)^2\expected{\Delta^2\hat n_p}_0\notag\\
&&+TT_p\eta_p(1-TT_p\eta_p)\expected{\hat n_p}_0\notag\\
&&-(TT_p\eta_p)^2\frac{\left(\expected{\hat n_p\hat n_a}_0-\expected{\hat n_p}_0\expected{\hat n_a}_0\right)^2}{\eta_a\expected{\Delta^2\hat n_a}_0+(1-\eta_a)\expected{\hat n_a}_0}.\notag\\
\eea

Error propagation can then be used to find the uncertainty in transmission estimation with the optimized intensity difference measurement, which is given by
\bea
\var{T}&=&\frac{\var{\hat n_-^{g_{\rm opt}}}}{\left|\frac{\partial\: T\eta_p\expected{\hat n_p}_r}{\partial T}\right|^2}\notag\\
&=&\frac{T}{\eta_p\expected{\hat n_p}_r}-\frac{T^2T_p}{\expected{\hat n_p}_r}H_a\left[1-\text{sech}(2s)\right]. \label{eq:IntensityDiffMeasure}
\eea
Comparing Eq.~\eqref{eq:IntensityDiffMeasure} with Eq.~\eqref{eq:LossyBTMSS}, it can be seen that the optimized intensity difference measurement saturates the bTMSS QCRB even in the presence of losses external to the system of interest.  In the derivation of Eq.~\eqref{eq:IntensityDiffMeasure}, we take into account that the partial derivative of $T\eta_p\expected{\hat n_p}_r$ with respect to $T$ is the same as the partial derivative on $\expected{\hat n_-^{g_{opt}}}$.  This is due to the auxiliary mode not probing the system and the optimal gain being a parameter that is set to minimize the noise around a given transmission level and does not change when performing a measurement.

\section{Conclusion}

In conclusion, we have shown that bright quantum states of light, such as the bTMSS and bSMSS, have a QFI per photon that approaches the ultimate bound for transmission estimation, given by the Fock state and the vTMSS, in the limit of high squeezing. As a result, given currently available squeezing levels of $s=2$, for a system with $T=99\%$ it would take only $\sim4.6$ ($\sim2.8$) times as many photons for a bTMSS (bSMSS) to obtain the same absolute precision as the Fock state while it would take 100 times more photons for a coherent state. Given that the bright quantum states we consider can be generated with a mean number of photons significantly higher that either the Fock state or vTMSS, they have a QCRB that can be many orders of magnitude lower while still offering a significant quantum advantage over a coherent state.

Additionally, we presented the effects of losses external to the system under study for both the probe  and auxiliary modes. External losses in the probe mode reduce the degree of quantum enhancement for all the states discussed.  As expected, the estimation functions of all states tend towards the one of a coherent state as external losses previous to the system increase given that the probing state approaches a coherent state in this limit.  For the bTMSS losses in the auxiliary mode also degrade the degree of quantum enhancement, with losses over 50\% causing the bTMSS estimation function to be higher than the one for the coherent state due to the additional loss of correlation between the probe and auxiliary modes.

Furthermore, we have shown that intensity measurements and optimal intensity difference measurements saturate the QCRB for the bSMSS and bTMSS, respectively, even in the presence of external losses. Such measurements can easily be implemented with current technology and thus offer a path for preforming transmission estimation at the QCRB. Given the range of applications for which transmission measurements are needed, the results presented pave the way to surpass the classical-state-of-the-art for real-life practical applications where there is a limit to the maximum power that can be used to probe a system.

This work was supported by the W.~M. Keck Foundation and by the National  Science  Foundation (NSF) (PHYS-1752938). C.~H-A. would like to acknowledge the support from FONDECYT Grant N$^{\circ}$ 11190078,  ANID-PAI grant 77180003, and Programa ICM Millennium Institute for Research in Optics (MIRO).

%\bibliography{QCRBReferences}

\appendix
\section{QFI for pure squeezed states}\label{app:PureTMSS}

Following the technique outlined in {\v{S}}afr{\'{a}}nek et al.~\cite{Safranek2015}, Gaussian states can be fully described by their displacement vector, $\vec{d}$, and covariance matrix, $\pmb{\sigma}$, which can be written in complex form as
\bea
\vec{d}&=&\Tr[\rho \vec{\hat A}]\\
\sigma_{i,j}&=&\Tr[\rho \{\Delta \hat A_i^{},\Delta \hat A_j^\dagger\}]\\
&=&2\Tr\left[\rho\left(\hat A_i^{}\hat A_j^\dagger+\hat A_j^\dagger\hat A_i^{}\right)\right]-2d_id_j,
\eea
where $\vec{\hat A}=(\hat a_1^{},\hat a_2^{},\cdots,\hat a_n^{},\hat a_1^\dagger,\hat a_2^\dagger,\cdots,\hat a_n^\dagger)^\text{T}$, $\Delta \hat A_i=\hat A_i-\langle\hat A_i\rangle$, $n$ is the number of modes, $\hat a^{}_i$ ($\hat a_i^\dagger$) is the annihilation (creation) operator for the $i^\text{th}$ mode, $\{\cdot,\cdot\}$ is the anticommutation relation, and $\rho$ is the density matrix. Here, we limit the discussion to the case of 2 modes, probe mode, $1\rightarrow p$, and auxiliary mode, $2\rightarrow a$. We work in the Heisenberg picture and take advantage of the following relations:
\bea
\hat S^\dagger_{p,a} \hat a^{}_p\hat S_{p,a}&=&\hat a^{}_p\cosh(s)-\hat a^\dagger_ae^{i\theta}\sinh(s)\\
\hat S^\dagger_{p,a} \hat a^{}_a\hat S_{p,a}&=&\hat a^{}_a\cosh(s)-\hat a^\dagger_pe^{i\theta}\sinh(s)\\
\hat B^\dagger \hat a^{}_p\hat B&=&\sqrt{T}\hat a^{}_p+\sqrt{1-T}\hat a^{}_\nu\\
\hat D^\dagger_p(\alpha) \hat a^{}_p\hat D^{}_p(\alpha)&=&\hat a^{}_p+\alpha\\
\hat D^\dagger_a(\beta) \hat a^{}_a\hat D^{}_a(\beta)&=&\hat a^{}_a+\beta,
\eea
where $\hat S_{p,a}$ is the two mode squeezing operator acting on the seed probe and auxiliary modes and $\hat D_{p}$ ($\hat D_{a}$) is the displacement operator for the probe (auxiliary) mode. For these calculation we used the notation introduced by Yuen for generating a bTMSS,  $\hat S_{p,a}\hat D_p(\alpha)\hat D_a(\beta)\ket{0,0}$, which matches the most common techniques of generating these states experimentally with a parametric amplifier.

With the use of the above relations, we find that for the TMSS the complex form of the covariance matrix takes the form
\begin{widetext}
\bea
\pmb{\sigma}&=&2\begin{pmatrix}
\Cov(\hat a_p^{},\hat a_p^\dagger)&\Cov(\hat a_p^{},\hat a_a^\dagger)&\Cov(\hat a_p^{},\hat a_p^{})&\Cov(\hat a_p^{},\hat a_a^{})\\
\Cov(\hat a_a^{},\hat a_p^\dagger)&\Cov(\hat a_a^{},\hat a_a^\dagger)&\Cov(\hat a_a^{},\hat a_p^{})&\Cov(\hat a_a^{},\hat a_a^{})\\
\Cov(\hat a_p^\dagger,\hat a_p^\dagger)&\Cov(\hat a_p^\dagger,\hat a_a^\dagger)&\Cov(\hat a_p^\dagger,\hat a_p^{})&\Cov(\hat a_p^\dagger,\hat a_a^{})\\
\Cov(\hat a_a^\dagger,\hat a_p^\dagger)&\Cov(\hat a_a^\dagger,\hat a_a^\dagger)&\Cov(\hat a_a^\dagger,\hat a_p^{})&\Cov(\hat a_a^\dagger,\hat a_a^{})
\end{pmatrix}\\
&=&\begin{pmatrix}
T\cosh(2s)+1-T&0&0&-\sqrt{T}e^{i\theta}\!\sinh(2s)\\
0&\cosh(2s)&-\sqrt{T}e^{i\theta}\!\sinh(2s)&0\\
0&-\sqrt{T}e^{\text-i\theta}\!\sinh(2s)&T\cosh(2s)+1-T&0\\
-\sqrt{T}e^{\text-i\theta}\!\sinh(2s)&0&0&\cosh(2s)
\end{pmatrix},
\eea
\end{widetext}
where $T$ is the transmission of the system, $s$ is the amplitude of the squeezing parameter for the TMSS, and $\theta$ defines the phase of the squeezing parameter. The diagonal terms are given by twice the covariance of the annihilation and creation operators of the probe mode and auxiliary mode, that is $\langle\hat a^\dagger_{i}\hat a^{}_i+\hat a^{}_{i}\hat a^\dagger_i\rangle-2\langle\hat a^\dagger_i\rangle\langle\hat a^{}_i\rangle$. The non-zero off diagonal terms are the cross terms of the covariance of the annihilation or creation operators of each mode, $2\langle\hat a^{}_{p}\hat a^{}_a\rangle-2\langle\hat a^{}_p\rangle\langle\hat a^{}_a\rangle$ for the upper right half or $2\langle\hat a^\dagger_{p}\hat a^\dagger_a\rangle-2\langle\hat a^\dagger_p\rangle\langle\hat a^\dagger_a\rangle$ for the lower left half.

Similarly, the displacement vector can be shown to take the form
\be
\vec{d}=\begin{pmatrix}
\sqrt{T}\left[\alpha\cosh(s)-\beta^*e^{i\theta}\sinh(s)\right]\\
\beta\cosh(s)-\alpha^*e^{i\theta}\sinh (s)\\
\sqrt{T}\left[\alpha^*\cosh(s)-\beta e^{\text -i\theta}\sinh(s)\right]\\
\beta^*\cosh(s)-\alpha e^{\text -i\theta}\sinh(s)
\end{pmatrix},
\ee
where $\alpha$ ($\beta$) is the complex field amplitude of the coherent state seeding of the probe (auxiliary) mode. As can be seen, $\vec{d}$ is given by the mean values of the creation and annihilation operators for each mode. The coherent state seed probe (auxiliary) has a mean photon number $|\alpha|^2$ ($|\beta|^2$) and a phase $arg(\alpha)$ ($arg(\beta)$).

The QFI for a two mode Gaussian state can then be calculated from the covariance matrix and displacement vector according to
\bea
F_Q(T)&=&\frac{1}{2\left(\left|\pmb\Sigma\right|-1\right)}\Bigg\{\left|\pmb\Sigma\right|\Tr\left[\left(\pmb\Sigma^{\text{-}1}\dot{\pmb\Sigma}\right)^2\right]\notag\\
&&+\sqrt{\left|\mathbb{I}+\pmb\Sigma^2\right|}\Tr\left[\left(\left(\mathbb{I}+\pmb\Sigma^2\right)^{\text{-}1}\dot{\pmb\Sigma}\right)^2\right]\notag\\
&&+4\left(\lambda_1^2-\lambda_2^2\right)\left(\frac{\dot{\lambda}_2^2}{\lambda_2^4-1}-\frac{\dot{\lambda}_1^2}{\lambda_1^4-1}\right)\Bigg\}\notag\\
&&+2\dot{\vec{d}}^\dagger\pmb{\sigma}^{\text{-}1}\dot{\vec{d}},\label{eqn:SafQFI}
\eea
where $|\cdot|$ is the determinant, $\mathbb{I}$ is the $4\times4$ identity matrix, $\pmb\Sigma=k.\pmb{\sigma}$ is the symplectic form of the covariance matrix, $k=diag(1,1,-1,-1)$, $\dot{\pmb\Sigma}$ is the element-wise derivative with respect to $T$, and $\lambda_i$ are the symplectic eigenvalues of $\pmb\Sigma$. For each symplectic eigenvalue $\lambda_i$ there exist another eigenvalue $-\lambda_i$, which is why equation~\eqref{eqn:SafQFI} only uses 2 eigenvalues. For our case, the positive symplectic eigenvalues for the TMSS can be calculated to be
\bea
\lambda_1&=&1\\
\lambda_2&=&T+(1-T)\cosh(2s).
\eea
Since $\lambda_1=1$ the derivative term, $\dot{\lambda}_1^2\,/\,(\lambda_1^4-1)$, is set to zero and in general a correction factor would need to be added. However, the correction factor is zero in our case.

Using Eq.~\eqref{eqn:SafQFI} we show the QFI takes the form
\begin{widetext}
\bea
F_Q^{TMSS}&=&\frac{\sinh^2(s)}{T-T^2}+\frac{|\alpha|^2\cosh^2(s)+|\beta|^2\sinh^2(s)-|\alpha||\beta|\cos(\Theta)\sinh(2s)}{T-T^2+T^2\text{sech}(2s)}\\
&=&\frac{\expected{\hat n_p}^\text{vac}}{T-T^2}+\frac{\expected{\hat n_p}^\text{bright}}{T-T^2+T^2\text{sech}(2s)},
\eea
\end{widetext}
where $\Theta=\theta-arg(\alpha)-arg(\beta)$ is the phase of the parametric process and sets which quadratures are squeezed or anti-squeezed. Thus, the phase of the QFI only depends on a single phase that is a linear combination of the squeezing phase and the phases of the input seeds, if both modes are seeded. As can be seen, the QFI is maximized for the double seeded case when $\cos(\Theta)=-1$, which corresponds to the phase that minimizes the intensity difference noise. The vacuum contribution in the QFI results form first three terms in Eq.~\eqref{eqn:SafQFI}, while the bright contribution in the QFI results from the fourth term that deals with displacement.

In the main text we consider the limit in which the stimulated or bright term, $\expected{\hat n_p}^\text{bright}$, dominates over the spontaneous one $\expected{\hat n_p}^\text{vac}$.  This happens when
\bea
\expected{\hat n_p}^\text{bright}&\gg&\frac{\left[1-T+T\text{sech}(2s)\right]}{1-T}\expected{\hat n_p}^\text{vac}\\
&\gg&\expected{\hat n_p}^\text{vac}+\frac{T\text{sech}(2s)}{1-T}\expected{\hat n_p}^\text{vac}\label{eqn:BrightLimit_init}
\eea
For $T<50\%$, the first term on the right hand side of Eq.~\eqref{eqn:BrightLimit_init} dominates and the bright limit is reached when there are more photon generated from seeding than spontaneously generated. For $T>90\%$ the second term can be an order of magnitude higher than the first and as $T\rightarrow100\%$, this term dominates and sets the bright limit. For large squeezing, taking into account that $\expected{\hat n_p}^\text{vac}=\sinh^2(s)$, we have that
\be
\lim_{s\rightarrow\infty}\text{sech}(2s)\sinh^2(s)\rightarrow\frac{1}{2},
\ee
a non-zero finite value such that the bright limit for transmissions near one can be approximated as
\be
\expected{\hat n_p}^\text{bright}\gg\frac{1}{2}\left(\frac{T}{1-T}\right).
\ee
This shows that the higher the transmission, the more seed probe photons are needed for the bright term to dominate.

The same approach can be used for the SMSS if we consider the auxiliary mode to be an uncorrelated coherent state. The lack of correlation between the states will make it such that the coherent state drops out of the calculations and does not contribute to the QFI. In this case the covariance matrix and displacement vector are
\bea
\pmb\sigma&=&\begin{pmatrix}
				T\cosh(2s)+1-T&0&-Te^{i\theta}\sinh(2s)&0\\
				0&1&0&0\\
				-Te^{\text -i\theta}\sinh(2s)&0&T\cosh(2s)+1-T&0\\
				0&0&0&1
				\end{pmatrix}\notag\\\\
\vec{d}&=&\begin{pmatrix}
		\sqrt{T}\left(\alpha\cosh(s)-\alpha^*e^{i\theta}\sinh(s)\right)\\
		\beta\\
		\sqrt{T}\left(\alpha^*\cosh(s)-\alpha e^{\text -i\theta}\sinh(s)\right)\\
		\beta^*
\end{pmatrix}.
\eea
If we again specialize to the contribution from the bright portion, the QFI for the bSMSS can be shown to take the form
\bea
F_Q(T)&=&2\dot{\vec{d}}^\dagger\pmb{\sigma}^{\text{-}1}\dot{\vec{d}}\\
&=&|\alpha|^2\frac{T+(1-T)\left[\cosh(2s)-\cos(\Theta)\sinh(2s)\right]}{T\left(1-2T(1-T)[1-\cosh(2s)]\right)},\notag\\
\eea
where $\Theta=\theta+2arg(\alpha)$. When the bSMSS is amplitude squeezed, $\cos(\Theta)=1$, the QFI is maximized  simplifies to
\be
F_Q(T)=\frac{\expected{n_p}_r}{T-T^2\left(1-e^{-2s}\right)},
\ee
with $\expected{n_p}_r=|\alpha|^2e^{-2s}$.

For both bright squeezed states, setting the squeezing parameter to zero returns the coherent state QFI. For the case of the bTMSS seeded with both a probe and auxiliary modes, the auxiliary coherent state does not contribute to the QFI, as was the case for the above calculations for the bSMSS, and the resulting QFI corresponds to the one of having a single coherent state probing the system.

\section{Effects of losses external to the system on the QFI}\label{app:Loss}

Given that Gaussian states remain Gaussian after any loss mechanism, the same approach as the one used to calculate the QFI in Appendix~\ref{app:PureTMSS} can be use to take into account external losses for the cases of the TMSS and SMSS. After taking into account the external losses on the probe and auxiliary modes, the TMSS covariance matrix takes the form
\begin{widetext}
\be
\sigma=\begin{pmatrix}
T_pT\eta_p\cosh(2s)+1-T_pT\eta_p\!&\!0&0\!&\!-\sqrt{T_pT\eta_p\eta_a}e^{i\theta}\!\sinh(2s)\\
0\!&\!\eta_a\cosh(2s)+1-\eta_a&-\sqrt{T_pT\eta_p\eta_a}e^{i\theta}\!\sinh(2s)\!&\!0\\
0\!&\!-\sqrt{T_pT\eta_p\eta_a}e^{-i\theta}\!\sinh(2s)&T_pT\eta_p\cosh(2s)+1-T_pT\eta_p\!&\!0\\
-\sqrt{T_pT\eta_p\eta_a}e^{-i\theta}\!\sinh(2s)\!&\!0&0\!&\!\eta_a\cosh(2s)+1-\eta_a
\end{pmatrix},
\ee
\end{widetext}
while the displacement vector becomes
\be
\vec{d}=\begin{pmatrix}
\sqrt{T_pT\eta_p}\left[\alpha\cosh(s)-\beta^*e^{i\theta}\sinh(s)\right]\\
\sqrt{\eta_a}\left[\beta\cosh(s)-\alpha^*e^{i\theta}\sinh(s)\right]\\
\sqrt{T_pT\eta_p}\left[\alpha^*\cosh(s)-\beta e^{-i\theta}\sinh(s)\right]\\
\sqrt{\eta_a}\left[\beta^*\cosh(s)-\alpha e^{-i\theta}\sinh(s)\right]
\end{pmatrix}.
\ee
The transmission of the additional beamsplitters used to model losses are multiplicative and as a result do not alter the functional form of the covariance matrix and displacement vector beyond the effect on the auxiliary mode, since there was no loss originally in that mode.

The symplectic eigenvalues, however, do change due to the additional losses and take the form
\bea
\lambda_1&=&\left(T_pT\eta_p-\eta_a\right)\sinh^2(s)+\Big\{1-\eta_a+T_pT\eta_p\left(2\eta_a-1\right)\notag\\&&+\left[\eta_a+T_pT\eta_p\left(1-2\eta_a\right)\right]\cosh(2s)\notag\\&&+\left(T_pT\eta_p-\eta_a\right)^2\sinh^4(s)\Big\}^{1/2}\\
\lambda_2&=&\left(T_pT\eta_p-\eta_a\right)\sinh^2(s)-\Big\{1-\eta_a+T_pT\eta_p\left(2\eta_a-1\right)\notag\\&&+\left[\eta_a+T_pT\eta_p\left(1-2\eta_a\right)\right]\cosh(2s)\notag\\&&+\left(T_pT\eta_p-\eta_a\right)^2\sinh^4(s)\Big\}^{1/2}.
\eea
As can be seen, the eigenvalues have additional terms due to the losses in the auxiliary mode. In the limit in which there are no losses on the auxiliary mode, the eigenvalues would be the same as for the pure state with $T\rightarrow T(T_p\eta_p)$.

The same approach can be used for the two Gaussian single-mode states we consider, the SMSS and the coherent state.  In the presence of losses, the SMSS will have a similar covariance matrix and displacement vector as the ones in Appendix~\ref{app:PureTMSS} but with $T\rightarrow T_pT\eta_p$. The QFI for the coherent state can once again be obtained from the QFI for the bTMSS or bSMSS by taking $s\rightarrow0$. Alternatively, it is possible to calculate the QFI for the coherent state directly, in which case only the fourth term on the right hand side of Eq.~\eqref{eqn:SafQFI} is non-zero. In this case the displacement vector would change to take the additional losses into account while its covariance matrix is always of the form $diag(1,1,1,1)$.

For the Fock state a different approach needs to be taken as it is not a Gaussian state. In this case, we can derive the lossy QCRB from its density matrix, which takes the form
\be
\rho_{Fock}^{lossy}=\sum_{k=0}^n\frac{n!}{k!(n-k)!}\left(T_pT\eta_p\right)^k\left(1\!-\!T_pT\eta_p\right)^{n-k}|k\rangle\!\langle k|,
\ee
where $n$ is the number of photons generated. The QFI can be calculated through the use of the following equation ~\cite{PARIS2009}
\be
F_Q^{Fock}=2\sum_{k,k'=0}^n\frac{\left|\langle k|\frac{\partial\rho_{Fock}^{lossy}}{\partial T}|k'\rangle\right|^2}{\rho_k+\rho_{k'}},
\ee
where $\rho_k=\langle k|\rho_{Fock}^{lossy}|k\rangle$.

\section{Optimal measurement for a single mode\label{app:Measure}}

As mentioned in the main text, intensity measurements, or photon counting for the Fock state, can saturate the QCRB for the considered single-mode states. The variance for an intensity measurement is given, after writing in normal ordering, by
\be
\var{\hat n}=\expected{\hat a^\dagger \hat a^\dagger \hat a\hat a+\hat a^\dagger \hat a}-\expected{\hat a^\dagger \hat a}^2.
\ee
Once losses, modeled as a BS, are taken into account, the intensity variance takes to form
\bea
\var{\hat n}&=&\expected{\hat B^\dagger \hat a^\dagger \hat a^\dagger \hat a\hat a+\hat a^\dagger \hat a\hat B}-\expected{\hat B^\dagger \hat a^\dagger \hat a\hat B}^2\\
&=&T^2\var{\hat n}_0+T(1-T)\expected{\hat n}_0,
\eea
where $\var{\hat n}_0$ and $\expected{\hat n}_0$ are the variance and mean before losses, respectively. As expected, as the transmission decreases the intensity variance tends towards the variance of a coherent state, $\var{\hat n}_{coherent}=\expected{\hat n}$.

For multiple sources of loss, the intensity variance takes the form
\bea
\var{\hat n}&=&T_2^2\var{\hat n}_{T_1}+T_2(1-T_2)\expected{\hat n}_{T_1}\\
&=&T_2^2\left(T_1^2\var{\hat n}_0+T_1(1-T_1)\expected{\hat n}_0\right)\notag\\&&+T_2(1-T_2)T_1\expected{\hat n}_{0}\\
&=&\left(T_2T_1\right)^2\var{\hat n}_0+T_2T_1\left(1-T_2T_1\right)\expected{\hat n}_0,\notag\label{eqn:BScommute}\\
\eea
where $T_1$ is the initial transmission followed by a transmission of $T_2$.  As expected, the variance takes the same form as the one for as a single BS with transmission $T=T_1T_2$ as the transmissions are multiplicative in this case.

To convert the variance in the intensity measurement to a variance in transmission measurement, we use the error propagation equation
\be
\var{T}=\frac{\var{\hat n_p}}{\left|\frac{\partial \expected{\hat n_p}}{\partial T}\right|^2},
\ee
with an intensity variance with external losses before and after the system
\be
\var{\hat n_p}=\left(T_pT\eta_p\right)^2\var{\hat n_p}_0+T_pT\eta_p(1-T_pT\eta_p)\expected{\hat n_p}_0,
\ee
and a mean photon number
\be
\expected{\hat n_p}=T_pT\eta_p\expected{\hat n_p}_0.
\ee
Thus, the variance in the estimation of the transmission from an intensity measurement takes the form
\be
\var{T}=\frac{\left(T_pT\eta_p\right)^2\!\expected{\!\Delta^2\hat n_p}_0\!\!+\!T_pT\eta_p(1\!\!-\!T_pT\eta_p)\!\expected{\hat n_p}_0}{\left|T_p\eta_p\expected{\hat n_p}_0\right|^2},
\ee
which can be rewritten as
\be
\expected{\hat n_p}_r\!\!\var{T}\!=\frac{T}{\eta_p}-T^2T_p\left(1-\frac{\var{\hat n_p}_0}{\expected{\hat n_p}_0}\right).\label{eq:lastequation}
\ee
As can be seen from Eq.~\eqref{eq:lastequation}, for any single-mode state intensity measurements result in  a transmission estimation enhancement with respect to a coherent state that depends on the Fano factor of the probing state. 

It is important to note that intensity measurements will not saturate the QCRB for transmission estimation for any single-mode state. For example, the vacuum single-mode squeezed state has a Fano factor that is twice that of a thermal state but in the limit of high transmissions its QCRB approaches the ultimate limit.

\end{document}